\documentclass{jetpl} \twocolumn

\lat

\title{Vector Area Theorem mapping in crystals and polarization stability of SIT\d solitons }

\rtitle{Vector Area Theorem mapping\ldots}

\sodtitle{Vector Area Theorem mapping in crystals and polarization
stability of SIT\d solitons }

\author{V.\,N.\,Lisin\thanks{e-mail: vlisin@kfti.knc.ru}}

\rauthor{V.\,N.\,Lisin}

\sodauthor{Lisin}

\address{Kazan Physical\d Technical Institute RAS,
420029 Kazan, Russia}

\dates{10 July 2001}{}

\abstract{The stability of polarization, areas, and number of
self\d induced transparency (SIT)\d solitons on an output from the
${LaF_3:Pr^{3+}}$ crystal is theoretically researched versus of
the polarization direction and the area of the input linearly
polarized laser pulse. For this purpose the Vector Area Theorem
is rederived and two\d dimensional Vector Area Theorem map is
obtained. The map is governed by the crystal symmetry and take
into account directions of the dipole matrix element vectors of
the  different site subgroups of optically excited ions. The
Vector Area Theorem mapping of the time evolution of the laser
pulse allows to highlight soliton polarization properties.}

\PACS{42.50.-p, 42.65.-k, 78.20.-e}

\begin{document}

\maketitle For an isotropic medium stability properties of SIT\d
solitons are determined by the Area Theorem.  The Area Theorem is
the name given to a theoretical result that governs the coherent
nonlinear transmission of short light pulses through isotropic
materials that have an absorption resonance very near the
frequency of the frequency of incident light, effectively
two-level media. In 1967 McCall and Hahn ~[1] identified a new
parameter (called "Area" and denoted by $\theta$) of optical
pulses travelling in such media, and then predicted that   obeys
the simple equation
\begin{equation}
\frac{\partial\theta}{\partial z} =- \frac{\alpha}2 \sin\theta.
\end{equation}
where $\alpha$ is the attenuation coefficient for the material.
The two most striking consequences of the Area Theorem are: (i)
pulses with special values of Area, namely all integer multiples
of $\pi$, are predicted to maintain the same Area during
propagation, and (ii) pulses with other values of Area must
change in propagation until their Area reaches one of the special
values. This property can be shown to be unstable for the odd
multiples, but the even multiples enjoy the full immunity of the
theorem. The Area Theorem ~(1) was derived for an isotropic
material in which the dipole matrix element vector of any ion is
parallel to the electrical field vector of the light pulse. By
contrast, the direction of the dipole matrix element vector of
any $Pr^{3+}$ ion in $LaF_3$ is not dependent from the electrical
field vector of the light pulse. So it is necessary to rederive
Area Theorem taking into account directions of the dipole matrix
element vectors of the different subgroups of $Pr^{3+}$ ions. The
$Pr^{3+}$ ions in a $LaF_3$ unit cell can replace $La^{3+}$ in six
different types of sites ($\pm\alpha,\pm\beta, \pm\gamma$). The
local environment of any of them has $C_2$-symmetry. The six local
$C_2$-symmetry axes are located in the plane normal to the $C_3$
axis and make the angle of $2\pi/6$ in this plane (Fig. 1). The
electrical dipole matrix element vector of the $Pr^{3+}$ ion
(optical transition $\Gamma_1\to\Gamma_1)$
\begin{equation}
{\bf p}_j = p  {\bf e}_j,\cdots j= 1,\cdots,6
\end{equation}
is directed ~[2] along of the local $C_2$-symmetry axis
\begin{equation}
{\bf e}_j=(\cos{(j\frac{2\pi}6)},\sin{(j\frac{2\pi}6)}),\quad
{\bf e}_z\cdot {\bf e}_j=0.
\end{equation}
where ${\bf e}_j$ is unit vector along the $C_{2j}$ axis. Here
axis $Z$ is directed along the $C_3$-axis and axis $X$ along
$\alpha$-axis.
    We define the Vector Area of light pulse as
\begin{equation}
\boldsymbol{ \Theta} = \frac p \hbar
\int_{-\infty}^{+\infty}dt{\bf E}(z,t),
\end{equation}
where ${\bf E}(z,t)$ is vector amplitude of a light pulse, $p$ is
electrical dipole matrix element, $\hbar$ is the Plank's constant.
Taking into account (2) and (3) and using arguments as Lamb ~[3]
we can write the Vector Area Theorem as follows:
\begin{equation}
\frac{\partial\boldsymbol{ \Theta}}{\partial z} =-\alpha\frac 1
6\sum\limits_{j=1}^{6} {\bf e}_j\sin(\boldsymbol{ \Theta}\cdot{\bf
e}_j),
\end{equation}
if a light pulse propagation is along $C_3$-axis. Here $\alpha$
is the linear attenuation coefficient for $LaF_3: Pr^{3+}$. It can
be seen from (5), as $\boldsymbol{ \Theta}\to 0$ , that
\begin{equation}
\frac{\partial\boldsymbol{ \Theta}}{\partial z}
=-\frac\alpha2\boldsymbol{ \Theta},
\end{equation}
as it must for a small pulse Area.

Equating the right part of the equation (5) to zero we can  find
special values of the Vector Area where $\partial\boldsymbol{
\Theta}/{\partial z}= 0$. But it to make much more obviously and
more easy from the graphical representation. We can rewrite eq (5)
as
\begin{equation}
\frac{\partial\boldsymbol{ \Theta}}{\partial z}
=\frac{\partial}{\partial \boldsymbol{ \Theta}}\frac \alpha
6\sum\limits_{j=1}^{6} \cos(\boldsymbol{ \Theta}\cdot{\bf e}_j),
\end{equation}
And the problem is reduced to a determination of points in a
two-dimensional plane, in which the function $\sum
\cos(\boldsymbol{ \Theta}\cdot{\bf e}_j)$ has extremes. The
circles and triangles in Fig.~2  give the contour plot of this
function. We easy find three types of special values of the
Vector Area, namely
\begin {equation}
{\boldsymbol \Theta}_c = m{\boldsymbol \Theta}_{+} + n{\boldsymbol
\Theta}_{-}, \end {equation}
\begin {equation}
{\boldsymbol \Theta}_u = {\boldsymbol \Theta}_c + {\bf u}_j,
\end{equation}
\begin {equation}
 {\boldsymbol \Theta}_s = {\boldsymbol \Theta}_c + {\bf
s}_j,
\end {equation}
are predicted to maintain the same Vector Area during
propagation. Here $m$ and $n$ are arbitrary integers and
\begin {equation}
{\boldsymbol \Theta}_{+} = \frac {2 \pi}{\cos(\pi/6)}{\bf
k}_1,\qquad {\boldsymbol \Theta}_{-} = \frac
{2\pi}{\cos(\pi/6)}{\bf k}_6;
\end {equation}
\begin {equation}
{\bf s}_j = \frac {\pi}{\cos( \pi/6)} {\bf k}_j,
\end {equation}
\begin {equation}
{\bf u}_j = \frac{\pi}{\cos^2(\pi/6)} {\bf e}_j,
\end {equation}
where the unit vectors ${\bf e}_j$ ~(3) and ${\bf k}_j$ are
directed along and between the $C_2$-axes accordingly:
\begin{equation}
{\bf k}_j=(\cos{(j\frac{2\pi}6- \frac{\pi}6)},\sin{(j\frac{2\pi}6-
\frac{\pi}6)})
\end{equation}
These peculiar points (8,9,10) give rise to a two-dimensional
lattice in a  $\boldsymbol \Theta$\d phase plane with basis
vectors ${\boldsymbol \Theta}_{+}$ and ${\boldsymbol
\Theta}_{-}$   ~(11) as you can see in Fig. 2. The unit cell of
the lattice is determined  by symmetry of the crystal. It is a
regular hexagon. The hexagon's centers are ${\boldsymbol
\Theta}_{c}$ ~(8) (centers of the circles in Fig. 2). As measured
from the hexagon center, coordinates of the six tops of the
hexagon are ${\bf u}_j$ ~(13) (centers of the triangles in Fig.
2), coordinates of middles of the sides of a hexagon are ${\bf
s}_j$ ~(12). As is easy to see from Eqs. (5,7) and definitions
(8,9,10), that in a neighborhood of these singular points the
Vector Area   behaves as
\begin{equation}
\frac{\partial{\boldsymbol \Theta}}{\partial z}
=-\frac\alpha2({\boldsymbol \Theta}-{\boldsymbol \Theta}_c),
\end{equation}
\
\begin{equation}
\frac{\partial\boldsymbol{ \Theta}}{\partial z}
=\frac\alpha4(\boldsymbol{ \Theta}-\boldsymbol{ \Theta}_u),
\end{equation}
\begin{equation}
\frac{\partial{\boldsymbol \Theta}}{\partial z}
=-\frac{\alpha}{6}(\boldsymbol{ \Theta}-{\boldsymbol \Theta}_s),
\end{equation}
if $\boldsymbol{ \Theta}-\boldsymbol{ \Theta}_s$   is directed
along the side of a hexagon, and
\begin{equation}
\frac{\partial\boldsymbol{ \Theta}}{\partial z}
=+\frac\alpha2(\boldsymbol{ \Theta}-\boldsymbol{ \Theta}_s),
\end{equation}
if $\boldsymbol{ \Theta}-\boldsymbol{ \Theta}_s$   is directed
perpendicularly to  the side of a hexagon. Therefore, for an
absorbing (amplifying) medium with $\alpha>0$ ($\alpha<0$), the
points (8) are type of a stable (unstable) knot, the points  (9)
are type of an unstable (stable) knot and the points  (10) are
type of a saddle in the  $\boldsymbol \Theta$\d phase plane. In
further we shall explore a case of the absorbing medium with
$\alpha>0$. If an input pulse Vector Area falls inside an unit
cell then the Vector Area must change in propagation until it
reaches the unit cell center. If an input Vector Area is not equal
to (9-10) and falls  on a side of a hexagon then the Vector Area
must change in a propagation until it reaches the middle of the
gexagon side. It is necessary to mark that the Vector Area
Theorem map (Fig. 2) allows easily to predict only the sum of the
pulse vector areas on an output from the sample. To determine the
number of the output SIT\d solitons and their polarizations and
areas we should solve the system of coupled Maxwell\d Bloch
equations.

 The input pulse, which Vector Area is
directed between the crystallographic axes and is equal, for
example, to ${\boldsymbol \Theta}_{+}$, excite only four
($\pm\alpha, \pm\gamma$ ) ion subgroups. It is $2\pi$\d pulse for
these ions. This pulse does not excite $\pm \beta$\d ions,
because ${\boldsymbol \Theta}_{+}\bot {\bf e}_{\pm \beta}$. To
the similarly previous case, the input pulse, which Vector Area
is equal to ${\boldsymbol \Theta}_{-}$, is $2\pi$\d pulse for
($\pm\alpha, \pm\beta$ )\d ion subgroups and does not excite
$\pm\gamma$\d ions. If the input Vector Area is parallel to the
${\boldsymbol \Theta}_{+}$ (${\boldsymbol \Theta}_{-}$) and fall
inside the unit cell ${\boldsymbol \Theta}_{c}=m{\boldsymbol
\Theta}_{+}$ (${\boldsymbol \Theta}_{c}=m{\boldsymbol
\Theta}_{-}$ ), then the time evolution of pulse, as it is easy
to show, may be described by the inverse scattering method. The
input pulse is splitting up on output, as for an isotropic
medium, into $m$ SIT- solitons, which Vector Area of any of them
is ${\boldsymbol \Theta}_{+}$ (${\boldsymbol \Theta}_{-}$). In
further we shall refer to these solitons as ${\boldsymbol
\Theta}_{+}$\d solitons and ${\boldsymbol \Theta}_{-}$\d
solitons. If input Vector Area is not parallel to the
${\boldsymbol \Theta}_{+}$ (${\boldsymbol \Theta}_{-}$) but fall
inside the unit cell ${\boldsymbol \Theta}_{c}=m{\boldsymbol
\Theta}_{+}$ (${\boldsymbol \Theta}_{c}=m{\boldsymbol
\Theta}_{-}$ ), then, as  show numerical calculations, input
pulse also is splitting up  into $m$ ${\boldsymbol \Theta}_{+}$
(${\boldsymbol \Theta}_{-}$)\d solitons on output.

If an input pulse Vector Area is directed along a
crystallographic axis, for example the axis $\alpha$, and is equal
to
\begin{equation}
{\boldsymbol \Theta}_{0}={\boldsymbol \Theta}_{+}+{\boldsymbol
\Theta}_{-},
\end{equation}
 then  all ($\pm\alpha, \pm\beta,\pm
\gamma$) ions are excited. The input pulse  is $2\pi$\d pulse for
($\pm\beta,\pm \gamma$ ) ion subgroups and $4p$\d pulse for
($\pm\alpha$ ) ions. The time evolution of the pulse is not
described by the inverse scattering method. The numerical
calculations have shown that if an input Vector Area ${\boldsymbol
\Theta}_{in}$  is parallel to ${\boldsymbol \Theta}_{0}$ and fall
inside the unit cell ${\boldsymbol \Theta}_{c}=m{\boldsymbol
\Theta}_{0}$, then the input pulse is splitting up into $m$
SIT-solitons, which a Vector Area of any of them is ${\boldsymbol
\Theta}_{0}$ . In further we shall refer to these solitons as
${\boldsymbol \Theta}_{0}$\d solitons. Let an input Vector Area is
not parallel to ${\boldsymbol \Theta}_{0}$ and fall inside the
unit cell ${\boldsymbol \Theta}_{c}=m{\boldsymbol \Theta}_{0}$.
Then, as you can see in Figs. ~2-3, a small change of an input
pulse polarization reduces to that each of the ${\boldsymbol
\Theta}_{0}$\d solitons is splitting up into ${\boldsymbol
\Theta}_{+}$\d  and ${\boldsymbol \Theta}_{-}$\d
solitons.Therefore a number of solitons and its polarization
strongly depend on a direction of a vector ${\boldsymbol
\Theta}_{in}$ concerning a crystallographic axis. This deduction
also is valid and in a  generally case when the input Vector Area
full down inside the unit cell ${\boldsymbol
\Theta}_{c}=m{\boldsymbol \Theta}_{+}+n{\boldsymbol \Theta}_{-}$,
where $m\not =n $. It is so because the unit cell center
coordinates may be  rewriting as ${\boldsymbol
\Theta}_{c}=(m-n){\boldsymbol \Theta}_{+}+n{\boldsymbol
\Theta}_{0}$ if $m > n$  or as ${\boldsymbol
\Theta}_{c}=(n-m){\boldsymbol \Theta}_{-}+m{\boldsymbol
\Theta}_{0}$  if  $n > m$. At first there are $(m -
n){\boldsymbol \Theta}_{+}$\d solitons for $m > n$, or $(n-
m){\boldsymbol \Theta}_{-}$\d solitons if $n > m$ on an output.
After that the number of solitons appearing on an output depends
on a direction of the vector
${\boldsymbol\Theta}_{in}-(m-n){\boldsymbol \Theta}_{+}$ or
${\boldsymbol \Theta}_{in}-(n-m){\boldsymbol \Theta}_{-}$
concerning a crystallographic axis. In a stable case the output
solitons are ${\boldsymbol \Theta}_{+}$\d solitons and
${\boldsymbol \Theta}_{-}$\d solitons and its number is $(m + n)$.

 For an amplifying medium  ${\boldsymbol \Theta}_{+}$\d
solitons and  ${\boldsymbol \Theta}_{-}$\d solitons are unstable
so the polarization of the output solitons must be directed along
crystallographic axis in a stable case.

It is necessary to mark that for circular polarization of laser
pulse the Area Theorem is (1) as in a case of isotropic medium.

To summarize, we have shown on an example of the model system
${LaF_3:Pr^{3+}}$, that the Vector Area  mapping of the pulse time
evolution during a propagation is effective method to analyze the
polarization properties of solitons.

We thank Ildar Ahmadullin for the help at assimilation of a
Fortran 90 and Ashat Basharov for the useful notes. The research
was supported by ISTC grant 737 and by the Russian Foundation for
Basic Research grant 00\ch 02\ch 16510.

\vfill\eject

Fig. 1. The directions of the local $C_2$-symmetry axes for the
different $ Pr^{3+}$ ion sites in the plane normal to the $C_3$-
axis of the ${LaF_3:Pr^{3+}}$ crystal.

Fig. 2.  Vector Area Theorem map. The $\boldsymbol{ \Theta}/2\pi$
projections  to axes $X$ and $Y$ are plotted on axes $X$ and
accordingly $Y$.  The vector ${\bf s}_1$ (12) and ${\bf u}_1$
(13) are shown in an upper of the figure. There are the basis
vectors ${\boldsymbol \Theta}_{+}$ and ${\boldsymbol
\Theta}_{-}$  ~(11) and the unit vectors along of the local
$C_2$-symmetry axes ($+\alpha,-\beta, -\gamma$) in a down. Vector
coordinates of some unit cell centers are also shown. Bold lines
$\bf0$ and $\bf1$ are mappings of the time evolution of laser
pulses which the input Vector Area is $mod({\boldsymbol
\Theta}_0)=4\pi$ and the angles between the directions of the
Vector Area and the crystallographic axis $\alpha$ are $0$ and
$-1$ degrees accordingly. In this case $\alpha L=20$, where $L$
is the sample length and $\alpha$ is the attenuation coefficient.
The bold line $\bf2$ is the mapping of the time evolution of the
laser pulse which the input Vector Area is $mod({2\boldsymbol
\Theta}_0)=8\pi$ and the angle between the directions of Vector
Area and the crystallographic axis $\alpha$ is $+1$ degree and
$\alpha L=40$. The bold line $\bf3$ is as $\bf2$ but $\alpha
L=80$.

Fig. 3.  The time evolution of the amplitude of the laser pulses
on the output of the sample. Values of the input Area, the angles
between the directions of the Vector Area and the crystallographic
axes $\alpha$ and the parameter $\alpha L$ for the curves
$\bf0$,$\bf1$ and $\bf2$ are the same as for the curves
$\bf0$,$\bf1$ and $\bf2$ in Fig. ~2. The parameter value $\alpha
L$ for the 0 degree curve in down of the figure is as for curve
$\bf 2$. The dot line is input pulse, $\tau_p$ is input pulse
duration. The Vector Area Theorem mapping (curves $\bf1$ and
$\bf2$ in Fig. 2)  allows easily to spot the polarizations and
the areas of the solitons in curves $\bf1$ and $\bf2$ in this
figure.


\begin{thebibliography}{99}
\bibitem{mccall}
S.\,L. McCall and E.\,L. Hahn, Phys.~Rev.~Lett. {\bf 18}, 908
(1967).

\bibitem{lisin}
V.\,N. Lisin, Pis'ma v ZhETF, {\bf 57}, 402 (1993) [JETP Lett.,
{\bf 57}, 415 (1993)].

\bibitem{lamb}
G.\,L. Lamb, in: {\sl Elements of Soliton Theory}, John Willey
and Sons, New York, 1980.
\end{thebibliography}
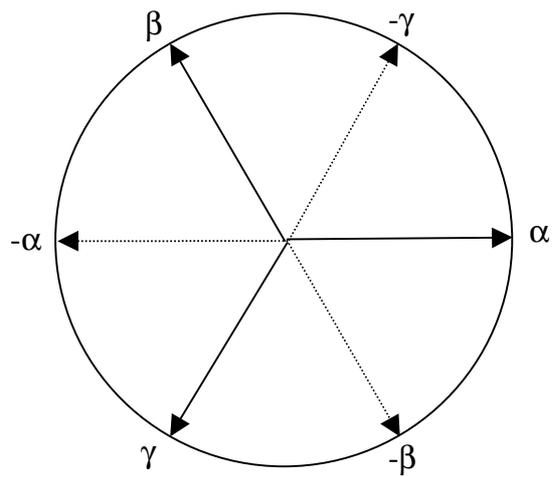
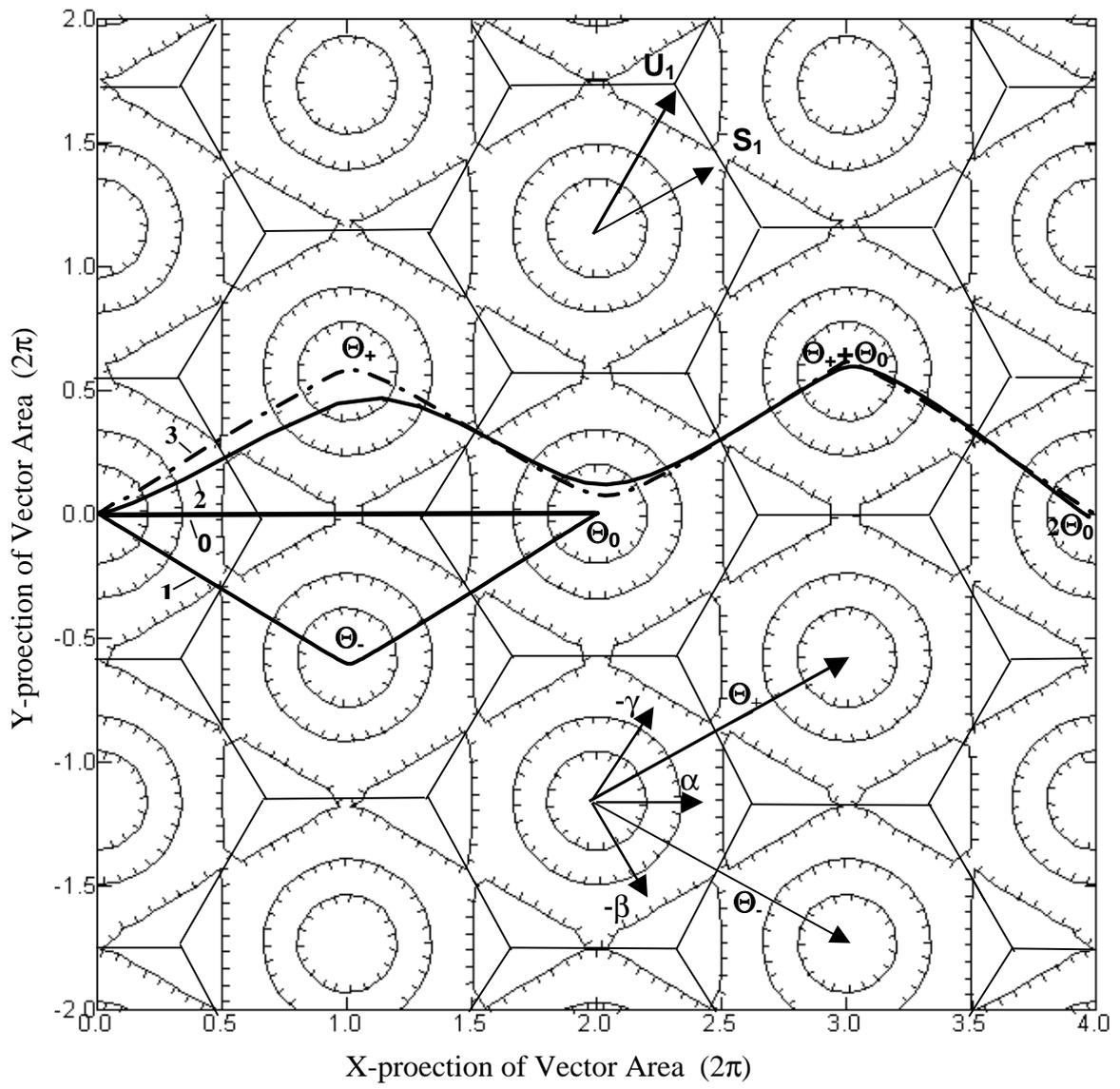
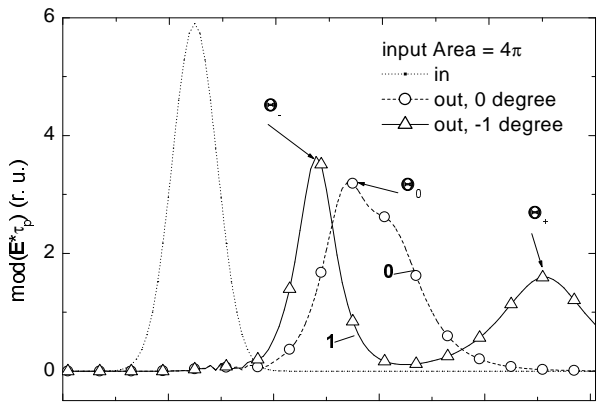
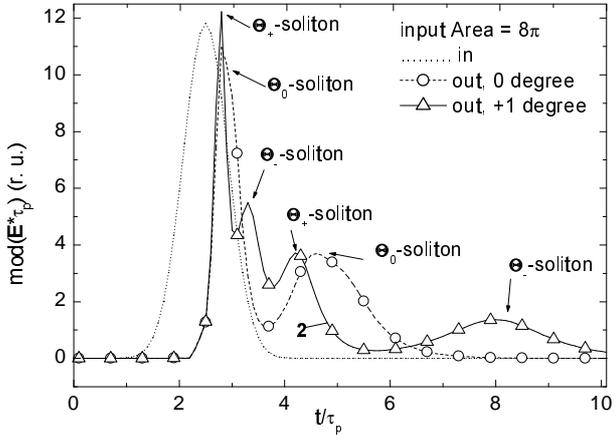
\end{document}